\documentstyle[11pt]{article}

\begin{document}
%

\begin{center}
{\large \bf The dark-matter world: Are there dark-matter galaxies?}

\vskip.5cm

W-Y. Pauchy Hwang\footnote{Correspondence Author;
 Email: wyhwang@phys.ntu.edu.tw; arXiv:xxxx (hep-ph, to be submitted)} \\
{\em Asia Pacific Organization for Cosmology and Particle Astrophysics, \\
Institute of Astrophysics, Center for Theoretical Sciences,\\
and Department of Physics, National Taiwan University,
     Taipei 106, Taiwan}
\vskip.2cm


{\small(October 10, 2011)}
\end{center}

\begin{abstract}
We attempt to answer whether neutrinos and antineutrinos, such as those in
the cosmic neutrino background, would clusterize among themselves or even
with other dark-matter particles, under certain time span, say
$1\, Gyr$. With neutrino masses in place, the similarity with the ordinary
matter increases and so is our confidence for neutrino clustering if time
is long enough. In particular, the clusterings could happen with some
seeds (cf. see the text for definition), the chance in the dark-matter world
to form dark-matter galaxies increases. If the dark-matter galaxies would
exist in a time span of $1\, Gyr$, then they might even dictate the
formation of the ordinary galaxies (i.e. the dark-matter galaxies get
formed first); thus, the implications for the structure of our Universe
would be tremendous.

\bigskip

{\parindent=0pt PACS Indices: 05.65.+b (Self-organized systems); 98.80.Bp
(Origin and formation of the Universe); 12.60.-i (Models beyond the standard
model); 12.10.-g (Unified field theories and models).}
\end{abstract}

\section{Introduction}

The phenomenon of clustering is a highly nonlinear effect. It
is against the tendency toward the uniform distribution or against
the democratization. Like the grown-up process, clustering seems to
acquire a life of certain form.

What comes to our minds is whether the dark-matter world, 25 \% of the
modern Universe (in comparison, only 5 \% in the ordinary matter), would
form the dark-matter galaxies, even before the ordinary-matter galaxies.
A dark-matter galaxy is by our definition a galaxy that does not see
any strong and electromagnetic interactions with our visible
ordinary-matter world. This fundamental question, though difficult, or
even impossible, to answer, deserves some thoughts, for the structural
formation of our Universe.

We wish to address ourselves about the likelihood of the
clustering - a question for which we could not begin with a theory
and prove the existence theorem of the clustering; rather, if there
is some clustering in some region, there is also de-clustering in the other
region(s). However, the large-scale clustering is so essential that all
life depends on it.

In the ordinary-matter world, strong and electromagnetic forces make the
clustering a different story - they manufacture atoms, molecules, complex
molecules, and chunks of matter, and then the stars and the galaxies; the
so-called "seeded clusterings". The seeds in the dark-matter world could 
be extra-heavy dark-matter particles - one such particle would be 
equivalent to thousands of ordinary-matter molecules. If neutrinos and 
antineutrinos be alone and don't interact with other dark-matter particles, 
the clustering would then be no-seed, and it might be too slow as compared 
to the time span of, e.g., $1\,Gyr$. So, we need to keep in mind whether 
a seeded clustering is possible - and it could be seeded in a different 
manner.

Maybe we could try out by other avenue - try to show that the dark-matter
world follows the rules of the ordinary-matter world, except the scales that
might be slightly different. Then we put together everything in the time
span of the age of our Universe. This is the first part of the "proof" which
we wish to follow in this note. The next part of the proof come from the fact
that atoms, molecules, complex molecules, and chunks of matter in fact
come from strong and electromagnetic forces - these seeds enhance the
clustering effect in a time span of $10^9\, Gyr$, the life of the early
Universe (the seeded clustering) - for the enhancement of this clustering
effect, weak interactions play little role, Of course, the gravitational
forces all add up the clustering effects because the effects, though tiny,
uniformly add up.

In the world that is full of unknowns - about 25\% of dark matter
and 70\% of dark energy, we are thinking of the world from the 5\%
visible ordinary-matter world. We know that the part in dark energy is
uniformly distributed. Is the dark-matter world clusterized? There
is some positive indication - such as the rotation curves of the spiral
galaxies.

We might argue that individual objects that have masses would have the
gravitational $1/ r^2$ force so that it leads to large-scale clustering,
provided that the time is long enough. Thus, evidence of neutrino's masses
is one positive indication for long-time gravitational large-scale
clustering. In fact, numerical simulations do give the evidence that
gravitational forces alone yield the clustering, if the time is sufficiently
long. But what we have in mind is that the sequence of atoms-molecules-complex
molecules-etc. yields the "seeds" of the clusterings - the seeded clustering
that might be relevant for the time span of, e.g., $1\, Gyr$, the age of the
early Universe. Such seeds for dark-matter galaxies may also come from heavy
dark-matter particles (with the mass greater than, e.g., $1\, TeV$). If
neutrinos, alone (but assumed that there are unknown not-so-weak interactions)
or together with other dark-matter species, could be found to aggregate at
large distances (somewhat larger than those for ordinary matter), it would
add the evidence of the seeded clustering. Note that the seeded clustering
could occur during the entire part of the early universe
and slightly longer, say, 1 Gyr or more. In the Big-Bang theory, neutrinos and
other weakly-interacting particles are manufactured earlier, than atoms
and molecules. The crucial question is whether the clustering in the
dark-matter world is seeded or no-seed.

We start with the minimal Standard Model of particles (which defines the
ordinary-matter world, as the zeroth-order approximation) and describe all
kinds of the known interactions - strong, electromagnetic, weak, and other
interactions. These particles do aggregate (into atoms, molecules, and then
macroscopically gravitational objects) under strong, electromagnetic, and
gravitational forces (in a time span of $1\, Gyr$, the life of the earlier
Universe) - for weak forces, they are much weaker, too weak for the
clustering problem that we are talking about. Presumably the same things
might have happened for the neutrinos and other dark-matter particles
except at a much larger scale. So, the first part of the
"proof" that neutrinos and other dark-matter particles are described
by the extended Standard Model does mean something important.

Thus, it is essential to understand the world of the extended Standard
Model (for ordinary-matter particles and dark-matter particles) for
the large-scale clustering of the dark-matter world. If everything turns
out to be similar to the particles in the "extended" Standard Model, we
would assert that there is at least the long-time large-scale clustering
of the dark-matter world - "long-time" in comparison to the age of our
Universe. Whether there is the "seeded clustering" (relevant in a time span
of $1\,Gyr$) is the next question to answer.

Note that the world of the minimal Standard Model, the visible ordinary-matter
world, seems to be extremely simple. One starts out with the electron, a
point-like spin-1/2 particle, and ends up with other point-like Dirac
particles, but with interactions through gauge fields modulated by the
Higgs fields. This is what we experimentally know, and it is a little strange
that it seems to be "complete" and that nothing else seems to exist. After
Dirac's equation we have searched for the point-like particles, now for eighty
years. It seems that Dirac equations explain all the relativistic point-like
particles and their interactions. So, why don't we formulate a working rule to
describe this fact?

Well, if the dark matter, occupying 25\% of the present-day Universe, would
clusterize or even form the dark-matter galaxies (particularly before the
ordinary-matter galaxies), the implications would be tremendous. That is why
we wish to "analyze" and to answer.

\section{Dirac Similarity Principle and Minimum Higgs Hypothesis}

In the minimal Standard Model that has been experimentally verified and that
describes the ordinary-matter world, it could be understood as a world consisting
of a set of point-like Dirac particles interacting through gauge fields modulated
by the Higgs fields. The only unknowns are neutrinos, which we believe may also
be point-llke Dirac particles. Thus, the minimal Standard Model is basically
a Dirac world interacting among themselves through gauge fields modulated by
the Higgs fields. In extending the minimal Standard Model, we try to keep the
"principle" of point-like Dirac particles intact - thinking of the eighty-year
experience some sort of sacred. On the other hand, the forty-year search for
Higgs (scalar fields) still in vain amounts to "minimal Higgs hypothesis".
This certainly offers us a very useful guide - the two working hypotheses
simplify a lot of things.

In other words, we follow another paper \cite{Hwang3} and introduce "the Dirac
similarity principle" that every "point-like" particle of spin-1/2 could be
observed in our space-time if it is "connected" with the electron, the
original spin-1/2 particle. For some reason\cite{Wu}, this clearly has
something to do with how relativity and the space-time structure gets married
with spin-1/2 particles. This is interesting since there are other ways to
write down (or, to express) spin-1/2 particles, but so far they are not seen
perhaps because they are not connected with the electron. In other words, the
partition between geometry (in numbers such as $4\times 4$ $\sigma/2$ in the
angular momentum) and space-time (such as $\vec r \times \vec p$) is similar
to the electron. We adopt "Dirac similarity principle" as the working
"principle" when we extend the Standard Model to include the
dark-matter particles as well.

These are "point-like" Dirac particles of which the size we believe is less than
$10^{-17}\,cm$, the current resolution of the length. Mathematically, the
"point-like" Dirac particles are described by "quantized Dirac fields" - maybe
via renormalizable lagrangian. The "quantized Dirac fields", which we can
axiomatize for its meaning, in fact does not contain anything characterizing
the size (maybe as it should be). In addition, the word "renormalization" may
contain something of the variable size.

Now that the dark matter are the species of the extended Standard
Model; then the dark matter world should exhibit the clustering effect,
just like the ordinary matter world. This is why we need to know to begin
with whether we could use the extended Standard Model to describe the
dark matter world.

So, if we could verify that the dark matter species are in the extension
of the Standard Model, then we have a proof of those guys do form clusters
just like the ordinary matter species.

In addition, we should pay another look at the Standard Model. After
forty years, the search for Higgs or scalar particle(s) is still in
vain, although the scalar fields are in some sense trivial (without spin,
etc. - no internal structure of any kind). So, the presence of these
particles should follow the "minimum Higgs hypothesis" - it might make a
lot of sense even though we don't know why our world looks like this.

Let's now "apply" the Dirac similarity principle and the minimum Higgs
hypothesis to our problem. The neutrinos
are now Dirac particles of some kind - so, right-handed neutrinos
exist and the masses could be written in terms of them. To make
Dirac neutrinos massive, we need a Higgs doublet for that. Is this
Higgs doublet a new Higgs doublet? In principle, we could use the
Standard-Model Higgs doublet and take and use the complex conjugate
(like in the case of quark masses) - the problem is the tininess of
these masses and if this would go it is definitely un-natural.

If a new and "remote" Higgs doublet would exist and the tininess of
the neutrino masses is explained by the neutrino couplings to the "remote" Higgs,
then it comes back to be "natural". Why are the neutrino couplings to "remote" Higgs
doublet should be small? - just similar to the CKM matrix elements (that is, the 31
matrix element is much small than the 21 matrix element); the other
"naturalness" reason.

We'd better look for the minimum number of Higgs multiplets
as our choice and the couplings to the "remote" Higgs would be much smaller
than to the ordinary one. As said earlier, this hypothesis makes the case of
the tiny neutrino masses very natural and, vice versa, we rephrase the natural
situation to get the hypothesis. Why do we adopt such hypothesis? For more than
forty years, we haven't found any solid signature for the Higgs; that the neutrinos
have tiny masses (by comparison with quarks and charged leptons) is another reason.

{\it Therefore, under "Dirac similarity principle" and "minimum Higgs hypothesis",
we have a unique Standard Model if the gauge group is determined or given.} These
two working hypotheses sort of summarize the characteristics of the (extended)
Standard Model.

That neutrinos have tiny masses can be taken as a signature that there is a heavy
extra $Z^{\prime 0}$, so that a new Higgs doublet should exist. This extra
$Z^{\prime 0}$ then requires the new "remote" Higgs doublet\cite{Hwang}.
This Higgs doublet also generates the tiny neutrino masses. This is one
possibility, $SU_c(3)\times SU_L(2)\times U(1) \times U(1)$; with minimum Higgs
hypothesis, we are talking about the unique extra $U(1)$ generation.

Alternately, we could require that the right-hand $SU_R(2)$ group exists
to restore the left-right symmetry\cite{Salam}. In this case, the left-handed
sector and the right-handed sector each has the Higgs doublet, each is the
left-right image of the other. The original picture\cite{Salam} contains many
options regarding the Higgs sector, but now the "minimum Higgs hypothesis"
makes the unique choice. In our terms, the right-handed Higgs would be the
"remote" Higgs for the left-handed species. That determines the size of the
coupling, including the tiny neutrino masses. Note that the only cut-off is
the masses of the right-handed gauge bosons (by keeping the left-right
symmetry). We believe that the phenomenology of the
said unique left-right symmetric model should be seriously pursued.

There is another option - the family option \cite{Family}. Here we are
curious at why there are three generations - is the family symmetry in fact
some sort of gauge symmetry because that the associated interactions are too
weak? We try to combine the Standard Model $SU_c(3)\times SU_L(2) \times U(1)$
with $SU_f(3)$, with $(\nu_{\tau R},\,\nu_{\mu R},\, \nu_{eR})$ the basic
$SU_f(3)$ triplet. Here $SU_f(3)$ has an orthogonal neutrino multiplet
since the right-handed neutrinos do not enter the minimal Standard Model.
In this way, we obtain the $SU_c(3) \times SU_L(2) \times U(1) \times SU_f(3)$
minimal model. Or, the right-handed indices could be removed altogether in the
family group, just like the other $SU_c(3)$ combining
with $SU_L(2)\times U(1)$ to avoid anomalies. Again, Dirac similarity
principle and minimum Higgs hypothesis saves the day - uniqueness in the
choice.

In this case \cite{Family}, the three family calls for $SU_f(3)$ and to make
the gauge bosons all massive the minimum choice would be a pair of Higgs
triplets - apparently a kind of broken gauge symmetry. Under the "minimum Higgs
hypothesis", the structure of the underlying Higgs mechanism is determined. Then,
neutrinos acquire their masses, to the leading order, with the aid of both the
Higgs triplets. Unless we could live with the unexplained duplications of
generations, the story might be the only way to go.

In other words, the particles in the ordinary-matter and dark-matter worlds
are divided into two categories: those in the dark-matter world and those
ordinary-matter particles as described by the minimal Standard Model, with
the only exception of neutrinos in the $SU_f(3)$-extended Standard Model
\cite{Family}. So, the neutrinos serve as the bridge between two worlds.
Otherwise, we can't think of the connections between the dark-matter and
ordinary-matter worlds and those spiral galaxies (such as our Milky Way)
are truly unthinkable.

All new species are "classified" as the so-called "dark matter", 25\%
of the present-day Universe (compared to 5\% of ordinary matter), in view
of their feeble (weak) interactions with the ordinary-matter species (in
the minimum Standard Model).

Let us focus on the extra $Z^{\prime 0}$ model to illustrate the
"minimum Higgs hypothesis" in some detail. There are left-handed neutrinos
belong to $SU_L(2)$ doublets while the right-handed neutrinos are singlets.
The term specified by
\begin{equation}
\varepsilon\cdot ({\bar\nu}_L,{\bar e}^-_L) \nu_R \varphi
\end{equation}
with $\varphi=(\varphi^0,\varphi^-)$ the new "remote" Higgs doublet could generate
the tiny mass for the neutrino and it is needed for the extra $Z^{\prime 0}$.

We need to introduce one working hypothesis on the couplings to the Higgs
- to the first (standard) Higgs doublet, from the electron to the top quark
we call it "normal" and $G_i$ is the coupling to the first Higgs doublet, and
to the second (extra, "remote") Higgs doublet the strength of the couplings for the Dirac
particles is down by the factor $(v/v')^2$ with $v$ the VEV for the standard Higgs
and $v'$ the VEV for the (remote) Higgs. Presumably, this contains in the
"minimum Higgs hypothesis". The hypothesis sounds very reasonable, similar
to the CKM matrix elements, and one may argue about the second power but for the
second Higgs fields some sort of scaling may apply.

With the working hypothesis, the coupling of the neutrinos to the standard
Higgs would vanish completely (i.e., it is natural) and its coupling to the
second (remote) Higgs would be $G_j (v/v')^2$ with $G_j$ the "normal" size.

The "minimum Higgs hypothesis" amounts to the assertion that there should be as
less Higgs fields as possible and the couplings would be ordering like the above
equation, Eq. (1).

Indeed, in the real world, neutrino masses are tiny
with the heaviest in the order of $0.1\, eV$. The electron, the lightest
Dirac particle except neutrinos, is $0.511\, MeV$\cite{PDG} or $5.11 \times 10^5\, eV$.
That is why the standard-model Higgs, which "explains" the masses of all other
Dirac particles, is likely not responsible for the tiny masses of the neutrinos.
The "minimum Higgs hypothesis" makes the hierarchy very natural.

In an early paper in 1987\cite{Hwang}, we studied the extra $Z^{\prime 0}$
extension paying specific attention to the Higgs sector - since in the Minimal
Standard model the standard Higgs doublet $\Phi$ has been used up by $(W^\pm,\,Z^0)$.
We worked out by adding one Higgs singlet (in the so-called 2+1 Higgs scenario) or
adding a Higgs doublet (the 2+2 Higgs scenario). It is the latter that we could add the
neutrino mass term naturally. (See Ref.\cite{Hwang} for details. Note that the
complex conjugate of the second "remote" Higgs doublet there is just the $\varphi$
above.)

The new Higgs potential follows the standard Higgs potential, except
that the parameters are chosen such that the masses of the new Higgs are much
bigger. The coupling between the two Higgs doublets should not be too big to
upset the nice fitting\cite{PDG} of the data to the Standard Model. All
these go with the smallness of the neutrino masses. Note that spontaneous symmetry
breaking happens such that the three components of the standard Higgs get absorbed
as the longitudinal components of the standard $W^\pm$ and $Z^0$.

As a parenthetical note, we could remark on the cancelation of the
flavor-changing scalar neutral quark currents. Suppose that we work with two
generations of quarks, and it is trivial to generalize to the physical case of
three. We should write
\begin{eqnarray}
({\bar u}_L,\,{\bar d}^\prime_L)d^\prime_R\Phi + c.c.;\nonumber\\
({\bar c}_L,\,{\bar s}^\prime_L)s^\prime_R\Phi + c.c.;\nonumber\\
({\bar u}_L,\,{\bar d}^\prime_L)u_R \Phi^*+c.c.;\nonumber\\
({\bar c}_L,\,{\bar s}^\prime_L)c_R \Phi^*+c.c.,
\end{eqnarray}
noting that we use the rotated right-handed down quarks and we also use the
complex conjugate of the standard Higgs doublet. This is a way to ensure that
the GIM mechanism\cite{GIM} is complete. Without anything to do the opposite, it
is reasonable to continue to assume the GIM mechanism.

There are additional questions such as: How about the couplings between quarks
(or charged leptons) and the (non-standard) remote Higgs? "Minimum Higgs
hypothesis" helps to set these couplings to zero or to be very small. For the new
gauge bosons (such as the right-handed $W^\pm,\,Z^0$), their large masses serve
as the cut-off. For the family gauge theory, all the couplings between family gauge
bosons and all Dirac particles, except neutrinos, vanish identically. We have a lot
to go in spelling out the details regarding the mass-generation (Higgs) mechanisms
- we eventually should pay a decent look at this mechanism rather than simply
attributing it the scalar fields.

To sum up, in any of the three extended standard models - the extra $Z^{\prime 0}$
\cite{Hwang}, the left-right model \cite{Salam}, and the family model
\cite{Family}, there is a clear distinction between the Standard-Model Higgs
and the remote Higgs - it is sufficient for quarks and charged leptons to couple
to the Standard-Model Higgs (to generate masses) and for neutrinos tiny masses
are generated with the aid of the remote Higgs. Anything that does not belong to the
minimal Standard Model could be classified as the "dark-matter particle".

\section{Symmetry missing or new}

Because the minimal Standard Model is a gauge theory (and is experimentally
proven), we could focus our discussion on the relations of these models with
symmetries - with the left-right model \cite{Salam} and the family model
\cite{Family} as some obvious examples. Contrary to this, the extra
$Z^{\prime 0}$ model is not symmetry-driven and will not be discussed it in
this section.

What we have done so far: We try to rephrase the Standard Model as a world of
point-like Dirac particles, or quantized Dirac fields, with interactions.
Dirac, in his relativistic construction of Dirac equations, was enormously
successful in describing the electron. (The point-like nature of the electron was
realized almost a century later.) Quarks, carrying other intrinsic degrees (color),
are described by Dirac equations and interact with the electron via gauge fields.
We also know muons and tau-ons, the other charged leptons. So, how about neutrinos?
Our first guess is also that neutrinos are point-like Dirac particles of some sort
(against Majorana or other Weyl fields). For some reasons, point-like Dirac
particles are implemented with some properties - that they know the other point-like
Dirac particles in our space-time. That is why we call it the "Dirac similarity
principle" to begin with. This is a world of point-like Dirac particles interacting
among themselves via gauge fields modulated by Higgs fields.

For our real minimal-Standard-model world, we begin with the electron and
end up with all three family of quarks and leptons, with gauge fields of
strong and electroweak interactions modulated by Higgs fields - the world
satisfied by the Dirac similarity principle and minimum Higgs hypothesis.
To proceed from there, we treat neutrinos as Dirac particles in accord with
the Dirac similarity principle.

As a matter of fact, we will treat only with renormalizable interactions
- with the spin-1/2 field power 3/2 and scalar and gauge fields power 1;
the total power counting less than 4. What is surprising is the role of
"renormalizability". We could construct quite a few such extensions of
the minimal Standard Model; they are all renormalizable - the present
extra $Z^{\prime 0}$, the left-right model (in the minimum sense), and
the recent proposed family gauge theory\cite{Family};
there are more. Apparently, we should not give up our original thinking
(such as the principle of renormalizability, the word of my mentor, late
Professor Henry Primakoff) though the road seems to have been blocked.

Our space-time is "defined" when the so-called "point-like Dirac particles"
are "defined", and vice versa. On the other hand, "point-like Dirac particles"
are in terms of "quantized Dirac fields". These concepts are "defined" together,
rather consistently.

We are amazed that a world of point-like Dirac particles, as
described by quantum field theory (the mathematical language), turns out
to be the physical world around us - that may define the space-time for us.
The interactions are mediated by gauge fields modulated slightly by Higgs
fields. There may be some new gauge fields, such as the extra $Z^{\prime 0}$,
or the missing right-handed partners\cite{Salam}, or the family gauge
symmetry\cite{Family}, or others.

There are two related remarks. The first remark related to the $SU_L(2)\times
SU_R(2) \times U(1)$ model\cite{Salam}. The second has to be related to the family (gauge)
symmetry\cite{Family}.

In the $SU_L(2)\times SU_R(2) \times U(1)$ model\cite{Salam}, suppose that in the
left and right parts each has one Higgs doublet (minimal) and we could try
to make the tiny neutrino masses in the right-handed sector - the "remote" Higgs
in the entire construction. Here we employ the so-called "minimum" working
hypothesis. This seems to be rather natural. We should think more about the
left-right symmetric model very seriously - except
that we should think of the Higgs mechanism in a real minimum fashion, judging
that we are looking for Higgs for about forty years. Thus, we have one left-handed
Higgs doublet and another right-handed (remote) Higgs doublet - due to spontaneous
symmetry breaking (SSB), only two neutral Higgs particles are left (after
spontaneous symmetry breaking). That means that
we are advocating a particular kind of the left-right symmetric model.

Regarding the family (gauge) symmetry, it is difficult to think
of the underlying reasons why there are three generations (of quarks and leptons).
But why? This is Raby's question; should we stop asking if the same question went
by without an answer for decades? This is why we promote the family symmetry
to the family gauge symmetry\cite{Family}. In both cases (left-right and family),
the proposed Dirac similarity principle and the "minimum Higgs hypothesis" both
may hold - an interesting and strange fact.

In both cases in the above remarks, it implies that, at temperature somewhat

higher than $1\,TeV$, there would be another phase transition - for the
spontaneous symmetry breaking. If most of the products from the phase transition
would remain to be dark matter, we would have most natural candidates for
the large amount (25\%) of dark matter.

In other words, those unseen particles so far, owing to their weak interactions with
ordinary matter, can be classified as "dark matter particles" in the extra
$Z^{\prime 0}$ model, or in the left-right model, or in the family gauge symmetry model.

The other interesting aspect is that the left-right symmetry is the missing
symmetry while the family gauge symmetry is the symmetry which we have found
but suspect that it is partially seen so far. It is difficult to speculate whether
the missing left-right symmetry would be found first or the family gauge symmetry
would first be seen. But it seems that the extra $Z^{\prime 0}$ picture has the least
chance in winning this game - no symmetry reason whatsoever. In mathematics, it turns
out that they all are self-consistent, some even "beautiful". In physics, 25\% of the
dark matter compared to only 5\% of ordinary matter means rooms in our imaginations -
even along the line of the extended Standard Model.

\section{The time span of $1\, Gyr$ -- the age of our early Universe}

Suppose that the spiral of the Milky Way is caused by the dark-matter aggregate of
four or five times the mass of the Milky Way, and similarly for other spiral galaxies.
This aspect will serve as a "basic fact" for our analysis of this section.

Let's look at our ordinary-matter world. Those quarks can aggregate in no time, to
hadrons, including nuclei, and the electrons serve to neutralize the charges also
in no time. Then atoms, molecules, complex molecules, and so on. These serve as
the seeds for the clusters, and then stars, and then galaxies, maybe in a time span
of $1\, Gyr$. The aggregation caused by strong and electromagnetic forces is fast
enough to give rise to galaxies in a time span of $1\, Gyr$. On the other hand, the
weak interactions proceed fairly slowly in this time span and they could not help
responsible for the galactic formation process as a whole.

On the other hand, the seeded clusterings might proceed with abundance of 
extra-heavy dark-matter particles such as familons and family Higgs, all 
greater than, e.g., $1\, TeV$. They belong to the dark-matter world, so they 
don't interact via strong or electromagnetic interactions (not directly, but 
indirectly through loops).  

The first part of the proof states that the ordinary-matter world and the dark-matter
world are jointly described by the extended Standard Model - proven by giving three
examples, the extra $Z^{\prime 0}$ model, the left-right symmetric model, and the
family $SU_3(3)$ gauge model, all renormalizable and obeying "Dirac similarity
principle" and the "minimum Higgs hypothesis". In other words, one, the last,
"extended Standard Model" would exist, to complete the saga of the "Standard Model"
for our space-time. Our Universe is after all consistent.

The statement that all ordinary-matter particles and dark-matter particles are described
by the extended Standard Model is important, but maybe not sufficient for the
clustering, and in particular for the clustering in a time span of, say, $1\, Gyr$.

In the minimum left-right symmetric model, the symmetry breaking for the right
sector happened much earlier, at the temperature greater than tens of $TeV$ (maybe at
least). Are there any remnants to begin manufacturing the clusters? The issue
is that all these are right-handed weak interactions but manufactured at very high
temperature ($>> 10 \,TeV$). The time span is $10^{-15}\, sec$ or shorter. For ordinary
left-handed weak interactions, it happens at $T\approx 0.3\, TeV$ but there is no
trace of the clustering effects (in a time span of $1\,Gyr$). We would conclude that
the clustering effects, relevant in a time span of $1\, Gyr$ (the age of the early
Universe), could not come from the manufacture of the right-handed sector.

On the other hand, it may be easy to analyze the minimal $SU_f(3)$ model - the strong
$SU_c(3)$ does give the aggregation effect that eventually gives rise to galaxies,
etc. So, $SU_f(3)$ is just another $SU(3)$ acting exclusively on the dark sector -
the related stuff serves as the seeds for the dark-matter clusters and even
galaxies - provided that the $SU_f(3)$ coupling is normal.

In fact, there are different scenarios, for example, the left-right symmetric model
may be recovered at $T\approx 3\, TeV$ (ten times the VEV at the electroweak scale,
maybe too low) and the family $SU_f(3)$ assumed to recovered at $T\approx 100\, TeV$.
Of course, the two models could be recovered at a different ordering. Here we assume
that the extra $Z^{\prime 0}$ model, the model without symmetry-driven, is less
favored.

In all cases, the seeded clustering effects, which may be relevant in a time span of
$1\,Gyr$, may come from the manufacture of the $SU_f(3)$ dark-matter sector, if the
$SU_f(3)$ coupling is normal (and strong); and the extra $Z^{\prime 0}$ model or the
left-right symmetric model seems to have nothing to do with the seeded clustering
effects.

\section{Are there dark-matter galaxies?}

In this note, we investigate the clustering of dark matter particles,
including neutrinos, by proposing to use the "extended" Standard Model to
describe dark matter particles. We extend the minimal Standard Model using
Dirac Similarity Principle and minimum Higgs hypothesis - the experience of
a half or a century. We do have a couple of very nice extended Standard Models,
inside which the clusterings would happen at different stages (judging from
the scales involved).

Using the extended Standard Model, we proceeded to look into the "seeded" 
clusterings, which may be relevant in a time span of about $1\,Gyr$, the 
life of the early Universe. The seeds might be the heavy dark-matter 
particles such as familons or family Higgs. In that case, dark-matter 
galaxies might be formed before ordinary-matter galaxies.  

If the dark-matter galaxies exist and play the hosts to the ordinary-matter
galaxies, the dark-matter hosts get formed at first. The picture of our
Universe is completely different from the conventional thinking, but it
makes more sense in terms of the 25\% dark matter versus the 5\% visible
ordinary matter. Apparently, the dark-matter galaxies, judging from the tiny
neutrino masses and the feeble interactions, would be much bigger, maybe by
a couple of orders (in length), than the visible ordinary-matter
galaxies to which they host.

\section*{Acknowledgments}
This research is supported in part by National Science Council project (NSC
99-2112-M-002-009-MY3).

\end{document}